\documentclass{article}
\usepackage{amsfonts}
\usepackage{amsmath}
\usepackage{makeidx}

\newtheorem{theorem}{Theorem}

\newtheorem{example}[theorem]{Example}

\input{tcilatex}

\begin{document}

\title{TI-games I: \\
An Exploration of Type Indeterminacy in Strategic Decision-making }
\author{Jerry Busemeyer\thanks{%
Indiana University, jbusemey@indiana.edu}, Ariane Lambert-Mogiliansky\thanks{%
Paris School of Economics, alambert@pse.ens.fr} \\
%EndAName
.}
\maketitle

\begin{abstract}
The Type Indeterminacy model is a theoretical framework that formalizes the
constructive preference perspective suggested by Kahneman and Tversky. In
this paper we explore an extention of the TI-model from simple to strategic
decision-making. A 2X2 game is investigated. We first show that in a
one-shot simultaneaous move setting the TI-model is equivalent to a standard
incomplete information model. We then let the game be preceded by a
cheap-talk promise exchange game. We show in an example that in the TI-model
the promise stage can have impact on next following behavior when the
standard classical model predicts no impact whatsoever. The TI approach
differs from other behavioral approaches in identifying the source of the
effect of cheap-talk promises in the intrinsic indeterminacy of the players'
type.

\ \ 

\end{abstract}

\date{}

\section{Introduction}

This paper belongs to a very recent and rapidly growing literature where
formal tools of Quantum Mechanics are proposed to explain a variety of
behavioral anomalies in social sciences and in psychology (see e.g. \cite%
{bus07,bus07',bus08,dan08,deutch99,franco07,franco08,khren07,lamura05,lamura08}%
). \medskip

The use of quantum formalism in game theory was initiated by Eisert et al. 
\cite{eisert99} who propose that models of quantum games can be used to
study how the extension of classical moves to quantum ones can affect the
analysis of a game.\footnote{%
From a game-theoretical point of view the approach consists in changing the
strategy spaces, and thus the interest of the results lies in the appeal of
these changes.} Another example is La Mura \cite{lamura03} who investigates
correlated equilibria with quantum signals in classical games. In this paper
we introduce some features of an extension of the Type Indeterminacy (TI)
model of decision-making \cite{almzz09} from simple decisions to strategic
decisions. The TI-model has been proposed as a theoretical framework for
modelling the KT(Kahneman--Tversky)--man, i.e., for the "constructive
preference perspective\textquotedblright .\footnote{%
\textquotedblleft \textit{There is a growing body of evidence that supports
an alternative conception according to which preferences are often
constructed -- not merely revealed -- in the elicitation process. These
constructions are contingent on the framing of the problem, the method of
elicitation, and the context of the choice\textquotedblright . }(Kahneman
and Tversky 2000).} Extending the TI-model to strategic decision-making is a
rather challenging task. Here we explore some central issues in an example
while the basic concepts and solutions are developed in a companion paper.
More precisely we investigate, in two different settings, a 2x2 game with
options, to cooperate and to defect and we refer to it as a Prisoner
Dilemma, PD\footnote{%
This is for convenience, as we shall see that the game is not perceived as a
true PD by all possible types of a player.}. In the first setting, the
players move simultaneously and the game is played once. In the second
setting, the simultaneous move PD game is preceded by a promise exchange
game. Our aim is to illustrate how the TI approach can provide an
explanation as to why cheap talk promises matter.\footnote{%
Cheap talk promises are promises that can be broken at no cost.} There
exists a substantial literature on cheap talk communication games (see for
instance \cite{Koessler08} for a survey). The approach in our paper does 
\textit{not} belong to the literature on communication games. The cheap talk
promise exchange stage is used to illustrate the possible impact of pre-play
interaction. Various behavioral theories have also been proposed to explain
the impact of cheap talk promises when standard theory predicts that there
is none. They most often rely on very specific assumptions amounting to
adding ad-hoc elements to the utility function (a moral cost for breaking
promises) or emotional communication \cite{frank88}. Our approach provides
an explanation relying on a fundamental structure of the model i.e., the
quantum indeterminacy of players' type. An advantage of our approach is that
the type indeterminacy hypothesis also explains a variety of other so called
behavioral anomalies such as framing effects, cognitive dissonance \cite%
{almzz09}, the disjunction effect \cite{bus07''} or the inverse fallacy \cite%
{franco08}. \medskip

A main interest with TI-game is that the Type Indeterminacy hypothesis can
modify quite significantly the way we think about games. Indeed, a major
implication of the TI-hypothesis is to extend the field of strategic
interactions. This is because actions impact not only on the payoffs but
also on the profile of types, i.e., on who the players \textit{are}. In a
TI-model, players do not have a well-determined (exogenously given) type.
Instead players' types change along the game together with the chosen
actions (which are modelled as the outcome of a measurement of the type). We
provide an example showing that an initially non-cooperative player can be
(on average) turned into a rather cooperative one by confronting him with a
tough player in a cheap talk promise exchange game.

Not surprisingly we find that there exists no distinction in terms of
predictions between the standard Bayes-Harsanyi and the Type Indeterminacy
approaches in a simultaneous move context. But in a multi-stage context
where the interaction at two different stages correspond to non-commuting
Game Situations\footnote{%
A Game Situation is an operator that measures the type of a player, see
below.} a move with no informational content or payoff relevance may still
impact on the outcome of the game.

\section{A TI-model of strategic decision-making}

In the TI-model a simple decision situation is represented by an \textit{%
observable\footnote{%
An observable is a linear operator.}} called a \textit{DS}. A decision-maker
is represented by his state or \textit{type}. A type is a vector $\left\vert
t_{i}\right\rangle $ in a\ Hilbert space. The measurement of the observable
corresponds to the act of choosing. Its outcome, the chosen item, actualizes
an \textit{eigentype\footnote{%
The eigentypes are the types associate with the eigenvalues of the
observable i.e., the possible outcomes of the measurement of the \textit{DS}%
. }} of the observable (or a \textit{superposition\footnote{%
A superposition is a linear combination of the form $\sum \lambda
_{i}\left\vert t_{i}\right\rangle ;\ \sum \lambda _{i}^{2}=1.\ $}} of
eigentypes if the measurement is coarse). It is information about the
preferences (type) of the agent. For instance consider a model where the
agent has preferences over sets of three items, i.e. he can rank any 3 items
from the most preferred to the least preferred. Any choice experiment
involving three items is associated with six eigentypes corresponding to the
six possible rankings of the items. If the agent chooses $a$ out of $\left\{
a,b,c\right\} $ his type is projected onto some superposition of the
rankings $\left[ a>b>c\right] $ and $\left[ a>c>b\right] .\ $The act of
choosing is modelled as a measurement of the (preference) type of the agent
and it impacts on the type i.e., it changes it (for a detailed exposition of
the TI-model see \cite{almzz09}). How does this simple scheme change when we
are dealing with strategic decision-making?\ \bigskip

We denote by \textit{GS}\ (for Game Situation)\textit{\ }an observable that
measures the type of an agent in a strategic situation, i.e., in a situation
where the outcome of the choice, in terms of the agent's utility, depends on
the choice of other agents as well. The interpretation of the outcome of the
measurement is that the chosen action is a \textit{best reply} against the
opponents' expected action\footnote{%
Note that a \textit{GS }is thus defined conditionnaly on the opponent's
type. In our companion paper we use the concept of \textit{GO} or Game
Operator, a complete collection of (commuting) \textit{GS} (each defined for
a specific opponent). The outcome of a \textit{GO} is an eigentype of the
game, it gives information about how a player plays against any possible
opponent in a specific game.}. This interpretation parallels the one in the
simple decision context. There, we interpret the chosen item as the \textit{%
preferred one} in accordance with an underlying assumption of rationality
i.e., the agent maximizes his utility (he chooses what he prefers). The
notion of revealed preferences (we shall use the term actualized rather than
revealed\footnote{%
The expression revealed preferences implicitely assumes that the prefrences
pre-existed the measurement and that they are uncovered by the measurement.
A central feature of the TI-model is precisely to depart from that
assumption. Preferences do not pre-exist the measurement. Preferences are in
a state of potentials that can be actualized by the measurement.}) and a
fortiori of actualized best-reply is problematic however. A main issue here
is that a best reply is a response to an \textit{expected} play. When the
expected play involves subjective beliefs there may be a problem as to the
measurability of the preferences. This is in particular so if subjective
beliefs are quantum properties.\footnote{%
If subjective beliefs and preferences are quantum properties that do not
commute then they cannot be measured simultaneously.} But in the present
context of maximal information games (see below for precise definition) we
are dealing objective probabilities so it is warranted to talk about
actualized best-reply.\medskip

TI-games are games with type indeterminate players, i.e., games
characterized by uncertainty. In particular, players do not know the payoff
of other players. The standard (classical) approach to incomplete
information in games is due to Harsanyi. It amounts to transforming the game
into a game of imperfect information where Nature moves at the beginning of
the game and selects, for each player, one among a multiplicity of possible
types (payoff functions). A player's own type is his private information.
But in a TI-game the players may not even know their \textit{own payoff}.
This is true even in TI-game of \textit{maximal information} where the
initial types are pure types.\footnote{%
Pure types provide maximal information about a player. But in a context of
indeterminacy, there is an irreducible uncertainty. It is impossible to know
all the type characteristics of a player with certainty. For a discussion
about pure and mixed types see Section 3.2 in \cite{dan08}.} Can the
Harsanyi approach be extended to TI-games? We shall argue that the
TI-paradigm gives new content to Harsanyi's approach. What is a fictitious
Nature's move in Harsanyi's setting becomes a real move (a measurement) with
substantial implications. And the theoretical multiplicity of types of a
player becomes a real multiplicity of "selves".\medskip

\paragraph{\textit{Types and eigentypes}}

We use the term \textit{type} to refer to the \textit{quantum pure state} of
a player. A pure type is maximal information about the player i.e., about
his payoff function. But because of (intrinsic) indeterminacy, the type is 
\textit{not} complete information\ about the payoff function in all games
simultaneously not even to the player himself (see \ref{dan08} for a
systematic investigation of non-classical indeterminacy with application to
social sciences).

In a TI-game we also speak about the \textit{eigentypes} of any specific
game $M$, these are \textit{complete information} about the payoff functions 
\textit{in a specific static game }$M$. Any eigentype of a player knows his
own $M$-game payoff function but he may not know that of the other players.
The eigentypes of a TI-game $M$\ are identified with their payoff function
in that game.

So we see that while the Harsanyi approach only uses a single concept, i.e.,
that of type and it is identified both with the payoff function and with the
player. In any specific TI-game $M$ we must distinguish between the type
which is identified with the player and the eigentypes (of $M$)\ which are
identified with the payoff functions in game $M$. A helpful analogy is with
multiple-selves models (see e.g., \cite{strotz} and \cite{fule06}). In
multiple-selves models, we are most often dealing with two "levels of
identity". These two levels are identified with short-run impulsive selves
on the one side and a long-run "rational self" on the other side. In our
context we have two levels as well: the level of the player (the type) and
the level of the selves (the eigentypes) which are to be viewed as potential
incarnations of the player \textit{in a specific game}.\textit{\medskip }

A central assumption that we make is that the reasoning leading to the
determination of the best-reply is performed at the level of the eigentypes
of the game. This key assumption deserves some clarification. What we do is
to propose that players are involved in some form of parallel reasoning: all
the active (with non-zero coefficient of superposition) eigentypes perform
their own strategic thinking. Another way to put it is that we assume that
the player is able to reason from different perspectives. Note that this is
not as demanding as it may at first appear. Indeed we are used in standard
game theory to the assumption that players are able to put themselves "in
the skin" of other players to think out how those will play in order to be
able to best-respond to that.\medskip

As in the basic TI-model, the outcome of the act of choosing, here a \textit{%
move,} is information about the (actualized) type of the player and the act
of choosing \textit{modifies} the type of the player e.g., from some initial
superposition it "collapses" onto a specific eigentype of the game under
consideration (see next section for concrete examples).

Finally, we assume that each player is an independent system i.e., there is
no entanglement between players.\footnote{%
In future research we intend to investigate the possibility of entanglement
between players.}\medskip\ 

We next investigate an example of a maximal information two-person game. The
objective is to introduce some basic features of TI-games in a simple
context and to illustrate an equivalence and some distinctions between the
Bayes-Harsanyi approach and the TI-approach.

\subsection{A single interaction}

Consider a 2X2 symmetric game, \textit{M}, and for concreteness we call the
two possible actions cooperate (C) and defect (D) (as in a Prisoner's
Dilemma game but as we shall see below for certain types, it is a
coordination game) and we define the preference types of game \textit{M}
also called the \textit{M-}eigentypes as follows:\medskip

$\theta _{1}:$ prefers to cooperate whatever he expects the opponent to do;

$\theta _{2}:$ prefers to cooperate if he expects the opponent to cooperate
with probability $p>q\ ($for some $q\leq 1)$ otherwise he prefers to defect;

$\theta _{3}:\ $prefers to defect whatever he expects the opponent to
do.\medskip \medskip

An example of these types is in the payoff matrices below where we depict
the row player's payoff:\medskip

$\theta _{1}:$ $%
\begin{pmatrix}
& C & D \\ 
C & 10 & 5 \\ 
D & 0 & 0%
\end{pmatrix}%
,\ \ \theta _{2}:$ $%
\begin{pmatrix}
& C & D \\ 
C & 10 & 0 \\ 
D & 6 & 8%
\end{pmatrix}%
,\ \ \theta _{2}:$ $%
\begin{pmatrix}
& C & D \\ 
C & 0 & 0 \\ 
D & 10 & 5%
\end{pmatrix}%
\medskip $

\bigskip

We shall now proceed to investigate this simultaneous move TI-game. We note
immediately that $\theta _{1}$ and $\theta _{3}$ are non-strategic while $%
\theta _{2}$ is, i.e., his best-reply will depend on what he expects the
opponent to do. The initial types are generally not eigentypes of the game
under consideration. Let player 1 be described by the superposition 
\begin{equation}
\left\vert t_{1}\right\rangle =\ \lambda _{1}\left\vert \theta
_{1}\right\rangle +\lambda _{2}\left\vert \theta _{2}\right\rangle +\lambda
_{3}\left\vert \theta _{3}\right\rangle ,\sum \lambda _{i}^{2}=1.\text{%
\footnote{%
As in the original TI-model, the coefficients of superposition are real
numbers and not complex numbers as in Quantum Mechanics. We motivation for
that can be found in \ref{almzz09}. }}  \label{player1}
\end{equation}%
We shall first be interested in the optimal play of player 1 when he
interacts with a player 2\ of different eigentypes. Suppose he interacts
with a player 2 of eigentype $\theta _{1}.$ Using the definitions of the
eigentypes $\theta _{i}\ $above and (\ref{player1}), we know by Born's rule%
\footnote{%
The calculus of probability in Quantum Mechanics is defined by Born's rule
according to which the probability for the different eigentypes is given by
the square of the coefficients of superposition.}\ that with probability $%
\lambda _{1}^{2}+\lambda _{2}^{2}$ player 1 plays $C\ ($because $\theta
_{2}^{{}}$'s best-reply to $\theta _{1}$ is \textit{C}$)\ $and he collapses
on the (superposed) type $\left\vert t_{1}^{\prime }\right\rangle =\frac{%
\lambda _{1}}{\sqrt{\lambda _{1}^{2}+\lambda _{2}^{2}}}\left\vert \theta
_{1}\right\rangle +\frac{\lambda _{2}}{\sqrt{\lambda _{1}^{2}+\lambda
_{2}^{2}}}\left\vert \theta _{2}\right\rangle .\ $With probability $\lambda
_{3}^{2}$ player 1 plays $D$ and collapses on the eigentype $\theta _{3}.\ $%
If instead player 1 interacts with a player 2 of type $\theta _{3}$\ then\
with probability $\lambda _{1}^{2}$ he plays $C$ and collapses on the
eigentype $\theta _{1}\ $and since $\theta _{2}^{\prime }s$ best-reply to $%
\theta _{3}$ is \textit{D,\ }with probability $\lambda _{2}^{2}+\lambda
_{3}^{2}$ he plays $D$ and collapses on type $\left\vert t"_{1}\right\rangle 
$=$\frac{\lambda _{2}}{\sqrt{\lambda _{3}^{2}+\lambda _{2}^{2}}}\left\vert
\theta _{2}\right\rangle +\frac{\lambda _{3}}{\sqrt{\lambda _{3}^{2}+\lambda
_{2}^{2}}}\left\vert \theta _{3}\right\rangle .\medskip $

We note that the probabilities for player 1's moves depends on the
opponent's type and corresponding expected play - as usual. More interesting
is that, as a consequence, the \textit{resulting type} of player 1 also
depends on the type of the opponent. This is because in a TI-model the act
of choice is a measurement that operates on the type and changes it. We
interpret the resulting type as the initial type modified by the
measurement. In a one-shot context, this is just an interpretation since
formally it cannot be distinguished from a classical informational
interpretation where the resulting type captures our revised beliefs about
player 1 (when our initial beliefs are given by (\ref{player1}).\medskip

We now consider a case when player 2's type is indeterminate as well: 
\begin{equation}
\left\vert t_{2}\right\rangle =\ \gamma _{1}\left\vert \theta
_{1}\right\rangle +\gamma _{2}\left\vert \theta _{2}\right\rangle +\gamma
_{3}\left\vert \theta _{3}\right\rangle ,\sum \gamma _{i}^{2}=1.
\label{player2}
\end{equation}

From the point of view of the eigentypes of a player (the $\theta _{i})$,
the situation can be analyzed as a standard situation of incomplete
information. We consider two examples:\medskip\ 

\begin{example}
Let $\lambda _{1}^{2}\geq q,$ implying that the eigentype $\theta _{2}$ of
player 2 cooperates and let $\gamma _{1}^{2}+\gamma _{2}^{2}\geq q\ $so\
the\ eigentype $\theta _{2}$ of player 1 cooperates as well.
\end{example}

\begin{example}
Let $\lambda _{1}^{2}\geq q$ so the eigentype $\theta _{2}$ of player 2
cooperates but now let $\gamma _{1}^{2}+\gamma _{2}^{2}<q\ $so here the\
eigentype $\theta _{2}$ of player 1 prefers to defect.
\end{example}

In Example 1 the types \textit{$\theta $}$_{1}$ and \textit{$\theta $}$_{2}\ 
$of both players pool to cooperate.\ So in particular player 1's resulting
type is a superposition of $\left\vert \theta _{1}\right\rangle $ and $%
\left\vert \theta _{2}\right\rangle $ with probability $\left( \lambda
_{1}^{2}+\lambda _{2}^{2}\right) $ and it is the eigentype $\left\vert
\theta _{3}\right\rangle $ with probability $\lambda _{3}^{2}.\ $In Example
2, player 1's eigentypes $\theta _{2}$\ and $\theta _{3\ }\ $pool to defect
so player 1's resulting type is a superposition of$\ \left\vert \theta
_{2}\right\rangle $ and $\left\vert \theta _{3}\right\rangle $ with
probability $\lambda _{2}^{2}+\lambda _{3}^{2}$ and $\left\vert \theta
_{1}\right\rangle $ with probability $\lambda _{1}^{2}.\ $So\ we see again
how the resulting type of player 1 varies with the initial (here superposed)
type of his opponent.\medskip

\textbf{Definition}

\textit{A pure static TI-equilibrium of a game }$M$\textit{\ with action set 
}$\mathit{A=}\left\{ a_{1},a_{2}\right\} $\textit{\ and strategy sets S}$%
_{1}=S_{2}=S\ $\textit{and initial\ types} $\left( \left\vert
t_{1}^{t=0}\right\rangle ,\left\vert t_{2}^{t=0}\right\rangle \right) $%
\textit{is}

\textit{i. A profile of pure strategies }$\left( s_{1}^{\ast },s_{2}^{\ast
}\right) \in S\times S\ $\textit{such that each one of the }$M-$\textit{%
eigentypes of each player maximizes his expected utility given the
(superposed) type of his opponent and the strategies played by the
opponent's eigentypes:}

\begin{equation*}
s_{1}^{\ast }\left( \theta _{iM}^{1}\right) =\arg \max_{s_{1}^{\prime }.\in
S_{{}}^{{}}}\sum_{\theta _{iM}^{2}}p\left( \left. \theta
_{iM}^{2}\right\vert \theta _{2}\right) u_{iM}\left( s_{1}^{\prime
},s_{2}^{^{\ast }}\left( \theta _{iM}^{2}\right) ,\left( \theta
_{iM}^{1},\theta _{iM}^{2}\right) \right) \ \text{\textit{for all} }\theta
_{iM}^{1}
\end{equation*}%
\textit{and similarly for player 2.}

\textit{ii. A corresponding profile of resulting types, one for each player
and each action:}

\begin{equation*}
\left\vert \left. t_{1}^{t=1}\right\vert a_{1}\right\rangle
=\sum_{_{iM};s_{1}^{\ast }\left( \theta _{iM}^{1}\right) =a_{1}}\frac{%
\lambda _{iM}}{\sqrt{\sum_{jM\neq iM}\lambda _{jM}^{2}\left( s_{1}^{\ast
}\left( \theta _{jM}^{1}\right) =a_{1}\right) }}\left\vert \theta
_{iM};s_{1}^{\ast }\left( \theta _{iM}^{1}\right) =a_{1}\right\rangle
\end{equation*}%
\textit{similarly for }$\left\vert \left. \theta _{1}^{t=1}\right\vert
a_{2}\right\rangle ,\ \left\vert \left. \theta _{2}^{t=1}\right\vert
a_{1}\right\rangle $\textit{\ and }$\left\vert \left. \theta
_{2}^{t=1}\right\vert a_{2}\right\rangle .\ $\medskip

For concreteness we shall now solve for the TI-equilibrium of this game in a
numerical example$.\ $Suppose the initial types are%
\begin{eqnarray}
\left\vert t_{1}\right\rangle &=&\ \sqrt{.7}\left\vert \theta
_{1}\right\rangle +\sqrt{.2}\left\vert \theta _{2}\right\rangle +\sqrt{.1}%
\left\vert \theta _{3}\right\rangle ,  \label{t1} \\
\left\vert t_{2}\right\rangle &=&\ \sqrt{.2}\left\vert \theta
_{1}\right\rangle +\sqrt{.6}\left\vert \theta _{2}\right\rangle +\sqrt{.2}%
\left\vert \theta _{3}\right\rangle .  \label{t2}
\end{eqnarray}

Given the payoff matrices above,\ the threshold probability $q$ that
rationalizes the play of \textit{C} for the eigentype $\theta _{2}\ $is $%
q=.666$. For the ease of presentation, we let $q=.7.\ $We know that the 
\textit{$\theta $}$_{2}$ of player 2 cooperates since $\lambda
_{1}^{2}=.7\geq q$ and so does the $\theta _{2}$ of player 1 since $\gamma
_{1}^{2}$+$\gamma _{2}^{2}=.8>q.\ $

In the TI-equilibrium of this game player 1 plays C with probability .9 and
collapses on $\left\vert t_{1}^{\prime }\right\rangle =\frac{\sqrt{.7}}{%
\sqrt{.7+.2}}\left\vert \theta _{1}\right\rangle +\frac{\sqrt{.2}}{\sqrt{%
.7+..2}}\left\vert \theta _{2}\right\rangle $ and with probability .1 player
1 plays $D$ and collapses on $\left\vert \theta _{3}\right\rangle .\ $Player
2\ plays $C$ with probability .8\ and collapses on $\left\vert t_{2}^{\prime
}\right\rangle =\frac{\sqrt{.4}}{\sqrt{.4+.4}}\left\vert \theta
_{1}\right\rangle +\frac{\sqrt{.4}}{\sqrt{.4+.4}}\left\vert \theta
_{2}\right\rangle $ and with probability .2, he plays $D$ and collapses on $%
\left\vert \theta _{3}\right\rangle .$

We note that the \textit{mixture actually played} by player 1 (.9C, .1D) is 
\textit{not} the best reply of any of his eigentypes. The same holds for
player 2. The eigentypes are the "real players" and they play pure
strategies. \medskip

We end this section with a comparison of the TI-game approach with the
standard incomplete information treatment of this game where the square of
the coefficients of superposition in (\ref{player1}) and (\ref{player2})\
are interpreted as players' beliefs about each other. The sole substantial
distinction is that in the Bayes-Harsanyi setting the players privately
learn their own type \textit{before} playing while in the TI-model they
learn it in the process of playing. A player is thus in the same
informational situation as his opponent with respect to his own play.
However under our assumption that all the reasoning is done by the
eigentypes, the classical approach and the TI-approach are
indistinguishable. They yield the same equilibrium outcome. The distinctions
are merely interpretational. \medskip

\textbf{Statement 1}

\textit{The equilibrium predictions TI-model of a simultaneous one-move game
are the same as those of the corresponding Bayes-Harsanyi model. \medskip }

A formal proof of Statement 1 can be found in our companion paper "TI-game
2".\medskip

This central equivalence result should be seen as an achievement which
provides support for the hypotheses that we make to extend the basic
TI-model to strategic decision-making. Indeed, we do want the non-classical
model to deliver the same outcome in a simultaneous one-move context.%
\footnote{%
We know that quantum indeterminacy cannot be distinguished from incomplete
information in the case of a single measurement. A simultaneous one-move
game corresponds to two single measurements performed on two non-entangled
systems.}\ We next move to a setting where one of the players is involved in
a sequence of moves. This is the simplest setting in which to introduce the
novelty brought about by the type indeterminacy hypothesis.

\subsection{A multi-stage TI-game}

In this section we introduce a new interaction involving player 1 and a
third player, a promise exchange game.\footnote{%
The reason for introducing a third player is that we want to avoid any form
of signaling. The exercise could be done with only two players but the
comparison between the classical and the TI-model would be less transparent.}
We assume that the \textit{GS}$\ $representing the promise game do not
commute with the \textit{GS} representing the game $M\ $(described in the
previous section$)$.\footnote{%
To each game we associate a collection of \textit{GS} each of which measures
the best reply a possible type of the opponent.} Player 1 and 3 play a
promise game where they choose between either making a non-binding promise
to cooperate with each other in game \textit{M} or withholding from making
such a promise. Our objective is to show that playing a promise exchange
game - with a third player - can increase the probability for cooperation
(decrease the probability for defection) between the player 1 and 2 in a
next following game \textit{M}. Such an impact of cheap-talk promises is
related to experimental evidence reported in Frank (1988) \bigskip

We shall compare two situations called respectively protocol I and II. In
protocol 1 player 1 and 2 play game \textit{M.} In protocol II we add a
third player, 3, and we have the following sequence of events:

\textit{step 1} Player player 1 and 3 play a promises exchange game $N$,
described below.

\textit{step 2} Player 1 and 2 play $M$.

\textit{step 3} Player 1 and 3 play $M$.\footnote{%
The reason why we have the interaction at \textit{step 3} is essentially to
motivate the promise exchange game. Our main interest will focus on the
interaction at \textit{step 2}.}

\medskip

\textit{The promise exchange game }

At \textit{step 1}, player 1 and 3 have to simultaneously select one of the
two announcements: "I promise to play cooperate", denoted, $P,\ $and "I do
not promise to play cooperate" denoted $no-P$. The promises are cheap-talk
i.e., breaking them in the next following games has no implications for the
payoffs i.e., at step 2\ or\ step 3. \medskip

There exists three eigentypes in the promise exchange game:

$\tau _{1}:\ $prefers to never make cheap-talk promises - let him be called
the "honest type";

$\tau _{2}:\ $prefers to make a promise to cooperate if he believes the
opponent cooperates with probability $p\geq q\ $(in which case he cooperates
whenever he is of type $\theta _{2}$ or $\theta _{1}$ or any superposition
of the 2$).\ $Otherwise he makes no promises - let him be called the
"sincere type";

$\tau _{\substack{ 3  \\ }}:\ $prefers to promise that he will cooperate
whatever he intends to do - he can be viewed as the "opportunistic
type".\bigskip

\textit{Information assumptions}

We make the following assumptions about players' information in the
multi-stage game:

i. All players know the statistical correlations (conditional probabilities)
between the eigentypes of the two (non-commuting) games.\footnote{%
So in particular they can compute the correlation between the \textit{plays}
in the different games.}

ii. At \textit{step 2,} player 2 knows that player 1 has interacted with
player 3 but he does not know the outcome of the interaction.

We note that ii. implies that we are not dealing with an issue of strategic
communication between player 1 and 2. No message is being received by player
2.

\subsubsection{\textit{The classical model}}

We first establish that in the classical setting we have the same outcome in
protocol I and at \textit{step 2} of protocol II. We already know from
Statement 1 that the predictions of a TI model of game \textit{M} are the
same as the prediction of the classical Bayes-Harsanyi model of the
corresponding incomplete information game.

We investigate in turn how the interaction between player 1 and 3 at \textit{%
step 1} affects the incentives and/or the information of player 1 and 2 at 
\textit{step 2}. Let us first consider the case of player 1. In a classical
setting, player 1 knows his own type, so he learns nothing from the promise
exchange stage. Moreover the announcement he makes is not payoff relevant to
his interaction with player 2. So the promise game has no direct implication
for his play with player 2. As to player 2, the question is whether he has
reason to update his beliefs about player 1. Initially he knows $\left\vert
t_{1}\right\rangle $ from which he derives his beliefs about player 1's
equilibrium play in game $M$. By our informational assumption (i) he also
knows the statistical correlations between the eigentypes of the two games
from which he can derive the expected play conditional on the choice at the
promise stage. He can write the probability of e.g., the play of \textit{D}
using the conditional probability formula: 
\begin{equation}
p\left( D\right) =p\left( P\right) p\left( \left. D\right\vert P\right)
+p\left( no-P\right) p\left( \left. D\right\vert no-P\right) .
\label{class1}
\end{equation}%
He knows that player 1 interacted with 3 but he does \textit{not} know the
outcome of the interaction. Therefore he has no new element from which to
update his information about player 1. We conclude that the introduction of
the interaction with player 3 at \textit{step 1} leaves the payoffs and the
information in the game \textit{M} unchanged. Hence, expected behavior at 
\textit{step 2} of protocol II is the same as in protocol I.\medskip

\subsubsection{\textit{The TI-model}}

Recall that the \textit{GS}$\ $representing the promise game do not commute
with the \textit{GS} representing the game $M$. We now write eq.(\ref%
{player1}) and (\ref{player2}) in terms of the eigentypes of game \textit{N, 
}i.e., of the promise stage eigentypes:%
\begin{equation*}
\left\vert t_{1}\right\rangle =\ \lambda _{1}^{\prime }\left\vert \tau
_{1}\right\rangle +\lambda _{2}^{\prime }\left\vert \tau _{2}\right\rangle
+\lambda _{3}^{\prime }\left\vert \tau _{3}\right\rangle \ \text{and\ }%
\left\vert t_{3}\right\rangle =\ \gamma _{1}^{\prime }\left\vert \tau
_{1}\right\rangle +\gamma _{2}^{\prime }\left\vert \tau _{2}\right\rangle
+\gamma _{3}^{\prime }\left\vert \tau _{3}\right\rangle .
\end{equation*}%
Each one of the $N-$eigentype can in turn be expressed in terms of the
eigentypes of game $M:$%
\begin{eqnarray}
\left\vert \tau _{1}\right\rangle &=&\ \delta _{11}\left\vert \theta
_{1}\right\rangle +\delta _{12}\left\vert \theta _{2}\right\rangle +\delta
_{13}\left\vert \theta _{3}\right\rangle  \label{RinT} \\
\left\vert \tau _{2}\right\rangle &=&\delta _{21}\left\vert \theta
_{1}\right\rangle +\delta _{22}\left\vert \theta _{2}\right\rangle +\delta
_{23}\left\vert \theta _{3}\right\rangle  \notag \\
\left\vert \tau _{3}\right\rangle &=&\delta _{31}\left\vert \theta
_{1}\right\rangle +\delta _{32}\left\vert \theta _{2}\right\rangle +\delta
_{33}\left\vert \theta _{3}\right\rangle  \notag
\end{eqnarray}%
where the $\delta _{ij}\ $are the elements of the basis transformation
matrix.\footnote{%
A basis transformation matrix links the eigentypes of the two \textit{GO }$%
\mathit{M\ }$and$\ N$ $:\medskip $%
\par
\textit{\ } $%
\begin{pmatrix}
\left\langle \tau _{1}\right\vert \left. \theta _{1}\right\rangle =\delta
_{11} & \left\langle \tau _{1}\right\vert \left. \theta _{2}\right\rangle
=\delta _{12} & \left\langle \tau _{1}\right\vert \left. \theta
_{3}\right\rangle =\delta _{13} \\ 
\left\langle \tau _{2}\right\vert \left. \theta _{1}\right\rangle =\delta
_{21} & \left\langle \tau _{2}\right\vert \left. \theta _{2}\right\rangle
=\delta _{22} & \left\langle \tau _{2}\right\vert \left. \theta
_{3}\right\rangle =\delta _{23} \\ 
\left\langle \tau _{3}\right\vert \left. \theta _{1}\right\rangle =\delta
_{31} & \left\langle \tau _{3}\right\vert \left. \theta _{2}\right\rangle
=\delta _{32} & \left\langle \tau _{3}\right\vert \left. \theta
_{3}\right\rangle =\delta _{33}%
\end{pmatrix}%
.$} Assume that player 3 is (initially) of type \textit{$\theta $}$_{3}$
with probability close to 1, we say he is a "tough" type$.\ $We\thinspace\
shall investigate the choice of between \textit{P} and \textit{no-P} of
player 1 i.e., the best response of the eigentypes $\tau _{i}$\ of player 1.

By definition of the $\tau _{i}$ type, we have that $\tau _{1}$ always plays 
\textit{no-P }and\ $\tau _{3}\ $always play \textit{P.\ \ }Now by
assumption, player 3 is of type $\theta _{3}$ who never cooperates.
Therefore, by the definition of $\tau _{2},$ player 1 of type $\tau _{2\ }$
chooses not to promise to cooperate, he plays $no-$\thinspace $P$\thinspace .

This means that at \textit{step 1} with probability $\lambda _{1}^{\prime
2}+\lambda _{2}^{\prime 2}$ player 1 plays $no-$\thinspace $P$\ and
collapses on $\left\vert \widehat{t}_{1}\right\rangle =\ \frac{\lambda
_{1}^{\prime }}{\sqrt{\left( \lambda _{1}^{2\prime }+\lambda _{2}^{2\prime
}\right) }}\left\vert \tau _{1}\right\rangle +\frac{\lambda _{2}^{\prime }}{%
\sqrt{\left( \lambda _{1}^{2\prime }+\lambda _{2}^{2\prime }\right) }}%
\left\vert \tau _{2}\right\rangle .$ \ With probability $\lambda
_{3}^{\prime 2}$ he collapses on $\left\vert \tau _{3}\right\rangle
.\medskip $

We shall next compare player 1's propensity to defect in protocol I with
that propensity in protocol II. For simplicity we shall assume the following
correlations: $\delta _{13}=\ \delta _{31}=$ 0, meaning that the honest type 
$\tau _{1},\ $never systematically defects and that the opportunistic guy $%
\tau _{3}\ $never systematically cooperates.\medskip

\textit{Player 1's propensity to defect in protocol I}

We shall consider the same numerical example as before i.e., given by (\ref%
{t1}) and\ (\ref{t2})\ so in particular we know that $\theta _{2}$ of player
1 cooperates so $p\left( D\left\vert \left\vert t_{1}\right\rangle \right.
\right) =\lambda _{3}^{2}$. But our objective in this section is to account
for the indeterminacy due to the fact that in protocol I the promise game is 
\textit{not} played. We have%
\begin{equation*}
\left\vert t_{1}\right\rangle =\ \lambda _{1}^{\prime }\left\vert \tau
_{1}\right\rangle +\lambda _{2}^{\prime }\left\vert \tau _{2}\right\rangle
+\lambda _{3}^{\prime }\left\vert \tau _{3}\right\rangle ,
\end{equation*}%
using the formulas in (\ref{RinT}) we substitute for the $\left\vert \tau
_{i}\right\rangle $ 
\begin{eqnarray*}
\left\vert t_{1}\right\rangle &=&\ \lambda _{1}^{\prime }\left( \delta
_{11}\left\vert \theta _{1}\right\rangle +\delta _{12}\left\vert \theta
_{2}\right\rangle +\delta _{13}\left\vert \theta _{3}\right\rangle \right)
+\lambda _{2}^{\prime }\left( \delta _{21}\left\vert \theta
_{1}\right\rangle +\delta _{22}\left\vert \theta _{2}\right\rangle +\delta
_{23}\left\vert \theta _{3}\right\rangle \right) \\
&&+\lambda _{3}^{\prime }\left( \delta _{31}\left\vert \theta
_{1}\right\rangle +\delta _{32}\left\vert \theta _{2}\right\rangle +\delta
_{33}\left\vert \theta _{3}\right\rangle \right) .
\end{eqnarray*}%
Collecting the terms we obtain%
\begin{eqnarray*}
\left\vert t_{1}\right\rangle &=&\ \left( \lambda _{1}^{\prime }\delta
_{11}+\lambda _{2}^{\prime }\delta _{21}+\lambda _{3}^{\prime }\delta
_{31}\right) \left\vert \theta _{1}\right\rangle +\left( \lambda
_{1}^{\prime }\delta _{12}+\lambda _{2}^{\prime }\delta _{22}+\lambda
_{3}^{\prime }\delta _{32}\right) \left\vert \theta _{2}\right\rangle + \\
&&\left( \lambda _{13}^{\prime }\delta +\lambda _{2}^{\prime }\delta
_{23}+\lambda _{3}^{\prime }\delta _{33}\right) \left\vert \theta
_{3}\right\rangle .
\end{eqnarray*}%
We know from the preceding section that both $\left\vert \theta
_{1}\right\rangle $ and $\left\vert \theta _{2}\right\rangle \ $choose to
cooperate so%
\begin{equation*}
p\left( D\left\vert \left\vert t_{1}\right\rangle \right. \right) =p\left(
\left\vert \theta _{3}\right\rangle \left\vert \left\vert t_{1}\right\rangle
\right. \right) .
\end{equation*}%
Using $\delta _{13}=$ 0$,\ $we obtain the probability for player 1's
defection in protocol I\textit{:}

\begin{equation}
p\left( D\left\vert \left\vert t_{1}\right\rangle \right. \right)
_{M}=\left( \lambda _{2}^{\prime }\delta _{23}^{{}}+\lambda _{3}^{\prime
}\delta _{33}^{{}}\right) ^{2}=\lambda _{2}^{2\prime }\delta
_{23}^{2}+\lambda _{3}^{2\prime }\delta _{33}^{2}+2\lambda _{2}^{\prime
}\delta _{23}\lambda _{3}^{\prime }\delta _{33}.  \label{MnoN}
\end{equation}%
\medskip

\textit{Player 1's propensity to defect in protocol II}

When the promise game is being played, i.e., the measurement \textit{N} is
performed, we can (as in the classical setting) use the conditional
probability formula to compute the probability for the play of $D$ 
\begin{equation}
p\left( D\left\vert \left\vert t_{1}\right\rangle \right. \right)
_{MN}=p\left( P\right) p\left( \left. D\right\vert P\right) +p\left(
no-P\right) p\left( \left. D\right\vert no-P\right) .  \label{M&N}
\end{equation}%
Let us consider the first term: $p\left( P\right) p\left( \left.
D\right\vert P\right) .$ We know that $p\left( P\right) =p\left( \left\vert
\tau _{3}\right\rangle \right) =\lambda _{3}^{2\prime }.$ We are now
interested in $p\left( \left. D\right\vert P\right) $ or $p\left(
D\left\vert \tau _{3}\right\rangle \right) .$\ $\left\vert \tau
_{3}\right\rangle $ writes as a superposition of the $\theta _{i}$ with $%
\theta _{1}$ who never defects, $\theta _{3}$ who always defect while $%
\theta _{2}^{\text{ }}$'s propensity to defect depends on what he expects
player 2 to do. We cannot take for granted that player 2 will play in
protocol II as he plays in protocol I.\ Instead we assume for now that
eigentype $\theta _{2\ }$ of player 2 chooses to cooperate (as in protocol
I) because he expects player 1's propensity to cooperate to be no less than
in protocol I. We below characterize the case when this expectation is
correct. Now if $\theta _{2}$ of player 2 chooses to cooperate so does $%
\theta _{2\ }$ of player $1$ and $p\left( D\left\vert \tau _{3}\right\rangle
\right) =\delta _{33}^{2}\ $so

\begin{equation*}
p\left( P\right) p\left( \left. D\right\vert P\right) =\lambda _{3}^{2\prime
}\delta _{33}^{2}
\end{equation*}%
We next consider the second term of (\ref{M&N}). The probability $p\left(
no-P\right) $ is $\left( \lambda _{1}^{2\prime }+\lambda _{2}^{2\prime
}\right) $ and the type of player 1 changes, he collapses on $\left\vert 
\widehat{t}_{1}\right\rangle =\frac{\lambda _{1}^{\prime }}{\sqrt{\left(
\lambda _{1}^{2\prime }+\lambda _{2}^{2\prime }\right) }}\left\vert \tau
_{1}\right\rangle +\frac{\lambda _{2}^{\prime }}{\sqrt{\left( \lambda
_{1}^{2\prime }+\lambda _{2}^{2\prime }\right) }}\left\vert \tau
_{2}\right\rangle $. Since we consider a case when $\theta _{2}$ of player 1
cooperates, the probability for defection of type $\left\vert \widehat{t}%
_{1}\right\rangle \ $is $\left( \frac{\lambda _{1}^{\prime }}{\sqrt{\left(
\lambda _{1}^{2\prime }+\lambda _{2}^{2\prime }\right) }^{{}}}\right)
^{2}\delta _{13}^{2}+\left( \frac{\lambda _{2}^{\prime }}{\sqrt{\left(
\lambda _{1}^{2\prime }+\lambda _{2}^{2\prime }\right) }}\right) ^{2}\delta
_{23}^{2}$. Recalling that $\delta _{13}^{{}}=0,$ we obtain that$\ p\left(
no-P\right) p\left( \left. D\right\vert no-P\right) $ is equal to

\begin{equation*}
\left( \lambda _{1}^{2\prime }+\lambda _{2}^{2\prime }\right) \left( \frac{%
\lambda _{2}^{\prime }}{\sqrt{\left( \lambda _{1}^{2\prime }+\lambda
_{2}^{2\prime }\right) }}\right) ^{2}\delta _{23}^{2}=\lambda _{2}^{2\prime
}\delta _{23}^{2}
\end{equation*}%
which gives 
\begin{equation}
p\left( D\left\vert \left\vert t_{1}\right\rangle \right. \right)
_{MN}=\lambda _{2}^{2\prime }\delta _{23}^{2}+\lambda _{3}^{2\prime }\delta
_{33}^{2}.  \label{MandN}
\end{equation}%
Comparing formulas in (\ref{MnoN})\ and\ (\ref{MandN}) :%
\begin{equation}
p\left( D\left\vert \left\vert t_{1}\right\rangle \right. \right)
_{MN}-p\left( D\left\vert \left\vert t_{1}\right\rangle \right. \right)
_{M}=-2\lambda _{2}^{\prime }\delta _{23}\lambda _{3}^{\prime }\delta _{33}
\label{Tuffplayer}
\end{equation}%
which can be negative or positive because the interference terms only
involves amplitudes of probability i.e., the square roots of probabilities.
The probability to play defect decreases (and thus the probability for
cooperation increases) when player 1 plays a promise stage whenever
\thinspace $2\lambda _{2}^{\prime }\delta _{23}\lambda _{3}^{\prime }\delta
_{33}<0.\ $In that case the expectations of player 2 are correct and we have
that the $\theta _{2}$ type of both players cooperate which we assumed in
our calculation above.\footnote{%
For the case the best reply of the $\theta _{2}$ types changes with the
performence of the promise game, the comparison between the two protocols is
less straightforward.} \medskip

\textbf{Result 1}: \textit{When player 1 meets a tough player 3 at step 1,
the probability for playing defect in the next following M game is not the
same as in the M game alone, }$p\left( D\left\vert \left\vert
t_{1}\right\rangle \right. \right) _{M}-p\left( D\left\vert \left\vert
t_{1}\right\rangle \right. \right) _{MN}\neq 0.\medskip $

It is interesting to note that $p\left( D\left\vert \left\vert
t_{1}\right\rangle \right. \right) _{MN}\ $is the same as in the classical
case. It can be obtained from the same conditional probability formula.

In order to better understand our \textit{Result 1,} we now consider a case
when player 1 meets with a "soft" player 3, i.e., a $\theta _{1}\ $type, at 
\textit{step 1}.\medskip

\emph{\textit{The \emph{s}}oft player 3 case}

In this section we show that if the promise stage is an interaction with a
soft player 3 there is no effect of the promise stage on player 1's
propensity to defect and thus no effect on the interaction at \textit{step 2}%
. \bigskip

Assume that player 3 is (initially) of type $\theta _{1}$ with probability
close to 1.$\ $What is the best reply of the \textit{N}-eigentypes of player
1, i.e., how do they choose between \textit{P} and \textit{no-P}? By
definition we have that $\tau _{1}$ always plays \textit{no-P }and\ \textit{$%
\tau $}$_{3}\ $always play \textit{P.\ \ }Now by the assumption we just made
player 3 is of type \textit{$\theta $}$_{1}$ who always cooperates so player
1 of type \textit{$\tau $}$_{2\ }$ chooses to promise to cooperate, he plays
\thinspace $P$\thinspace .

This means that at \textit{t=1} with probability $\lambda _{1}^{\prime 2}$
he collapses on $\left\vert \tau _{1}\right\rangle \ $and with probability $%
\lambda _{2}^{\prime 2}+\lambda _{3}^{\prime 2}$ player 1 plays $P$\ and
collapses on $\left\vert \widehat{t}_{1}\right\rangle =\ \frac{\lambda
_{2}^{\prime }}{\sqrt{\lambda _{2}^{2\prime }+\lambda _{3}^{2\prime }}}%
\left\vert \tau _{2}\right\rangle +\frac{\lambda _{3}^{\prime }}{\sqrt{%
\lambda _{2}^{2\prime }+\lambda _{3}^{2\prime }}}\left\vert \tau
_{3}\right\rangle .\ $We shall compute the probability to defect of that
type.\footnote{%
Recall that $\tau _{1}$ never defects.} We write the type vector $\left\vert 
\widehat{t}_{1}\right\rangle $ in terms of the \textit{M}-eigentypes,%
\begin{eqnarray*}
\left\vert \widehat{t}_{1}\right\rangle &=&\left( \frac{\lambda _{2}^{\prime
}}{\sqrt{\lambda _{2}^{\prime 2}+\lambda _{3}^{\prime 2}}}\right)
^{{}}\left( \delta _{21}\left\vert \theta _{1}\right\rangle +\delta
_{22}\left\vert \theta _{2}\right\rangle +\delta _{23}\left\vert \theta
_{3}\right\rangle \right) \\
&&+\left( \frac{\lambda _{3}^{\prime }}{\sqrt{\lambda _{2}^{\prime
2}+\lambda \prime _{3}^{2}}}\right) \left( \delta _{31}\left\vert \theta
_{1}\right\rangle +\delta _{32}\left\vert \theta _{2}\right\rangle +\delta
_{33}\left\vert \theta _{3}\right\rangle \right)
\end{eqnarray*}%
As we investigate player 1's \textit{M}-eigentypes' best reply, we again
have to make an assumption about player 2's expectation. And the assumption
we make is that he believes that player 1's propensity to defect is
unchanged, so as in protocol I the $\theta _{2}$ of both players cooperate
and only $\theta _{3}$ defects. We have 
\begin{equation*}
p\left( D\left\vert \left\vert \widehat{t}_{1}\right\rangle \right. \right)
_{MN}=\left[ \frac{\lambda _{2}^{\prime }}{\sqrt{\lambda _{2}^{\prime
2}+\lambda _{3}^{\prime 2}}}\delta _{23}+\frac{\lambda _{3}^{\prime }}{\sqrt{%
\lambda _{2}^{\prime 2}+\lambda _{3}^{\prime 2}}}\delta _{33}\right] ^{2}
\end{equation*}%
\begin{equation*}
p\left( D\left\vert \left\vert \widehat{t}_{1}\right\rangle \right. \right)
_{MN}=\frac{1}{\lambda _{2}^{2\prime }+\lambda _{3}^{2\prime }}\left[
\lambda \prime _{2}^{2}\delta _{23}^{2}+\lambda \prime _{3}^{2}\delta
_{33}^{2}+2\lambda _{2}^{\prime }\lambda _{3}^{\prime }\delta _{23}\delta
_{33}\right]
\end{equation*}%
The probability for defection is thus%
\begin{equation*}
p\left( D\left\vert \left\vert t_{1}\right\rangle \right. \right)
_{MN}=P\left( \tau _{1}\right) p\left( D\left\vert \left\vert \tau
_{1}\right\rangle \right. \right) +P\left( \widehat{t}_{1}\right) p\left(
D\left\vert \left\vert \widehat{t}_{1}\right\rangle \right. \right) =
\end{equation*}%
\begin{equation*}
0+\left( \lambda _{2}^{2\prime }+\lambda _{3}^{2\prime }\right) \frac{1}{%
\lambda _{2}^{2\prime }+\lambda _{3}^{2\prime }}\left[ \lambda _{2}^{2\prime
}\delta _{23}^{2}+\lambda _{\text{3}}^{2\prime }\delta {}_{33}^{2}+2\lambda
_{2}^{\prime }\lambda _{3}^{\prime }\delta _{23}\delta _{33}\right] =\lambda
_{2}^{2\prime }\delta _{23}^{2}+\lambda _{\text{3}}^{2\prime }\delta
{}_{33}^{2}+2\lambda _{2}^{\prime }\lambda _{3}^{\prime }\delta _{23}\delta
_{33}.
\end{equation*}%
Comparing with eq. (\ref{MnoN}) of protocol I we see that here 
\begin{equation*}
p\left( D\left\vert \left\vert t_{1}\right\rangle \right. \right)
_{M}=p\left( D\left\vert \left\vert t_{1}\right\rangle \right. \right) _{MN}
\end{equation*}%
There is NO effect of the promise stage. This is because the interference
terms are still present. We note also that player 2 was correct in his
expectation about player 1's propensity to defect. \medskip

\textbf{Result 2}

\textit{If player 1's move at step 1 does not separate between the
N-eigentypes that would otherwise interfere in the determination of his play
of D at step 2 then }$p\left( D\left\vert \left\vert t_{1}\right\rangle
\right. \right) _{M}=p\left( D\left\vert \left\vert t_{1}\right\rangle
\right. \right) _{MN}.$\textit{\ } $\medskip $

Let us try to provide an intuition for our two results. In the absence of a
promise stage (protocol I) both the sincere and opportunistic type coexist
in the mind of player 1. Both these two types have a positive propensity to
defect. When they coexist they interfere positively(negatively) to
reinforce(weaken) player 1's propensity to defect. When playing the promise
exchange game the two types may either separate or not. They separate in the
case of a tough player 3. Player 1 collapses either on a superposition of
the honest and sincere type (and chooses \textit{no-P}) or on the
opportunistic type (and chooses \textit{P}). Since the sincere and the
opportunistic types are separated (by the first measurement, game $N$) there
is no more interference. In the case of a soft player 3 case, the play of
the promise game does not separate the sincere from the opportunistic guy,
they both prefer $P.\ $As a consequence the two \textit{N}eigentypes
interfere in the determination of outcome of the next following \textit{M}
game as they do in protocol I. \bigskip

In this example we demonstrated that in a TI-model of strategic interaction,
a promise stage does make a difference for players' behavior in the next
following performance of game \textit{M.\ }The promise stage makes a
difference because it may destroy interference effects that are present in
protocol I.\medskip

Quite remarkably the distinction between the predictions of the classical
and the TI-game only appears in the \textit{absence} of the play of a
promise stage (with a tough player). Indeed the probability formula that
applies in the TI-model for the case the agent undergoes the promise stage (%
\ref{MandN}) is the same as the conditional probability formula that applies
in the standard classical setting.\medskip

\textit{The cheap-talk promise paradox}

When promises that have no commitment or informational value affect
behavior, we may speak about a cheap talk paradox (with respect to
established theory). In particular we may have the case that despite the
fact that all types pool to make cheap-talk promises (we only have
non-revealing equilibria), they nevertheless affect subsequent play. Our
paper does not exactly address this case. This is because on the one hand
playing the promise game always separates between the $\tau _{1}$ and $\tau
_{3}.$\ On the other hand the promises are not communicated to player 2.
Yet, because the analysis focuses on the separation between $\tau _{2}$ and $%
\tau _{3}$ (and by its information assumption avoids Bayesian updating with
respect to $\tau _{1}),\ $it suggests two possible explanations for why
cheap talk promises may matter:

1. Unobserved separation

Here the idea is that the promise game actually does trigger separation
between types (like in protocol 2 with a taught type$)$. Reaching the
promise response is more difficult for the reciprocating type $\tau _{2}$
than for the opportunistic $\tau _{3}.$ It takes longer time to do the
reasoning. The act of playing breaks the indeterminacy of player 1 but that
is not observed by player 2, both $\tau _{2}$ and $\tau _{3}\ $choose to
promise or they choose differently but player 2 does not learn about it. In
that case the TI-model's predictions in the next following PD game with or
without a prior promise exchange game are not the same. We have an impact of
cheap-talk promises.

2. Observed pooling

The second line of explanation of the paradox follows a different logic. It
relies on the observation that if the observer has the classical model in
mind, his predictions are incorrect. When he confronts his predictions in
protocol 2 (which are the same as his predictions in protocol 1) with the
actual outcome of protocol 2, he notes a difference. This is because simply
he did not account for the interference effects. So here the explanation is
not that pooling in cheap-talk promises changes behavior but that there is
an error in the modeling of the pooling outcome.

\subsubsection{Possible fields of application of TI-games}

We have learned from this explorative example that TI-games may bring forth
new results in the context of multi-stage game or when a game is preceded by
some form of "pre-play". We conjecture that the Type Indeterminacy approach
may bring new light on the following issues:

- Players' choice of selection principle in multiple equilibria situation;

In Camerer \ref{Cam03} the author reports about experiments where a pre-play
auction impacts on the principle of selection among multiple equilibria in a
coordination game. The pre-play auction for the right to play the
coordination game tended to push toward the payoff-dominant equilibrium
compared with the no pre-play case. In a TI-game, preferences with respect
to the equilibrium selection criteria can be modified by pre-play.

- The selection of a reference point;

According to experiments (see ) playing a contest before an ultimatum game
can affect the equilibrium offer and acceptance threshold. In a TI-game the
pre-play of a contest may change the preferences of the players with respect
to what they feel entitled to in an ultimatum game played next.

- The sunk cost fallacy;

According to numerous experiments and casual evidence people seem to be the
victims of the sunk cost fallacy. In an experiment, people who were offered
a year subscription to the theater showed (on average) a greater propensity
to go to the theater than people who were not offered subscription. In a
TI-game the pre-play decision to purchase a subscription may modify people's
valuation of theater plays.

- Path-dependency;

A single (little probable) move which radically modifies the type of a
player can yield significative implications for the path of future play.

\section{Concluding remarks}

In this paper we have explored an extension of the Type Indeterminacy model
of decision-making to strategic decision-making in a maximal information
context. We did that by means of an example of a 2X2 game that we
investigate in two different settings. In the first setting the game is
played directly. In the second setting the game is preceded by a promise
exchange game. We first find that in a one-shot setting the predictions of
the TI-model are the same as those of the corresponding Bayes-Harsanyi game
of incomplete information. This is no longer true in the multiple move
setting. We give an example of circumstances under which the predictions of
the two models are not the same. We show that the TI-model can provide an
explanation for why a cheap-talk promises matter. The promise game may
separate between potential eigentypes and thereby destroy interference
effects that otherwise contribute to the determination of, e.g., the
propensity to defect in the next following game.

Last we want to emphasize the very explorative character of this paper. A
companion paper TI-game 2 develops the basic concepts and solutions of
TI-games. We believe that this avenue of research has a rich potential to
explain a variety of puzzles in (sequential) interactive situations and to
give new impulses to game theory.

\end{document}